\documentclass[twocolumn,showpacs,prb,aps,amsmath,amssymb]{revtex4-1}
\usepackage{graphicx}
\usepackage{amsmath}
\usepackage{amssymb}
\usepackage{color}
\usepackage[T1]{fontenc}
\newcommand{\nup}[1]{\hat{n}_{#1 \sigma}}
\newcommand{\D}[1]{\hat{D}_{#1} }
\newcommand{\muup}[1]{\mu_{#1 \sigma}}
\newcommand{\ciup}[1]{\hat{c}_{#1 \sigma}}
\newcommand{\cdgiup}[1]{{\hat{c}^\dag}_{#1 \sigma}}
\newcommand{\mudw}[1]{\mu_{#1 \bar{\sigma}}}
\newcommand{\ndw}[1]{\hat{n}_{#1 \bar{\sigma}}}
\newcommand{\nupzr}[1]{{\langle \hat{n}_{#1 \sigma} \rangle}_0}
\newcommand{\ndwzr}[1]{{\langle \hat{n}_{#1 \bar{\sigma}} \rangle}_0}
\newcommand{\qup}[1]{q_{#1 \sigma}}

\begin{document}
\title{Electrical field induced shift of the Mott Metal-Insulator transition in thin films}
\author{D. Nasr Esfahani, L. Covaci and F. M. Peeters}
\affiliation{Departement Fysica, Universiteit Antwerpen,
Groenenborgerlaan 171, B-2020 Antwerpen, Belgium}

\begin{abstract}
The ground state properties of a paramagnetic Mott insulator are investigated in the presence of 
an external electrical field using the inhomogeneous Gutzwiller approximation for a single band Hubbard model
in a slab geometry. The metal insulator transition is shifted towards higher Hubbard repulsions by applying an electric field perpendicular to the slab. The spatial distribution of site dependent quasiparticle weight shows that the quasiparticle weight is maximum
in few layers beneath the surface. Moreover only at higher Hubbard repulsion, larger than the bulk critical U, the electric field will be totally screened only for centeral cites. Our results show that by presence of an electric field perpendicular to a thin film made of
a strongly correlated material, states near the surface will remain metallic while the bulk becomes insulating after some critical U. In contrast, in the absence of the electric field the surface becomes insulating before the bulk.
\end{abstract}

\pacs{71.30.+h, 71.27.+a, 73.61.-r} 

\maketitle
\section{Introduction}
The rich physics of strongly correlated materials in combination with the need to overcome the scaling limits of current silicon based semi-conductor materials
in microelectronic industry has resulted in an increased activity in this field. Special attention has been focused on vanadium dioxide($VO_2$) which shows 
an abrupt Metal Insulator Transition(MIT) near room temperature due to a structural phase transition.\cite{cavalleri} 
One has found that an electrical field is able to trigger 
MIT in $VO_2$, without any structural transition, which is mostly dominated by electron correlations rather than a Peierles distortion \cite{kim}. Also, a first order MIT is observed by applying an electrical field in a two terminal model of $VO_2$ \cite{ruzmetov,caviglial} without
electrical breakdown of the material. Note that an, electric field driven MIT and metal-superconductor transitions have been observed at the interface between $LaAlO_3$ and $SrTiO_3$. \cite{caviglial} These kind of transitions may be related to a charge transfer mechanism. A nonlinear dependence of the conductivity on the electrical field is reported for the highly correlated transition metal chalcogenide $ Ni{(͑S,Se)}_2 $ and a continuous MIT is observed in this case \cite{husmann}.

In this paper we investigate the behavior of the ground state of a single band Hubbard model\cite{hubbard} in the presence of a perpendicular electric field by using the Gutzwiller approximation (GA)\cite{gutz}. Originally GA is rooted in the Gutzwiller wave function used to reduce the contribution of high energy states due to Hubbard repulsion and it was shown to be exact in the limit of infinite dimensions.\cite{MV1,MV2,geb} 

While  an analytical solution exists only for one dimension \cite{lieb}, in comparison to other approximate methods the GA is equivalent to a slave boson mean field approach(SBMF)\cite{kotliar} for zero temperature but in contrast to dynamical mean field theory (DMFT)\cite{georges}, it is not able to give any information about higher and lower Hubbard bands. Instead it gives a reasonable understanding about the low energy excitations near the Fermi surface\cite{bunemann} by supplying the quasiparticle weight of electrons such that one is then able to describe the mobility of electrons. Also, GA cannot give any information about the insulating state, instead we are only able to investigate 
the properties of the system by approaching the transition point, $U_c$, from below \cite{geb}.
This method was used by Brinkman-Rice\cite{rice} to investigate the MIT of the single band Hubbard model 
and it allowed them to predict the critical Hubbard repulsion which is finite in two and three dimensions($U_c=16t$ for 3D). While not as accurate as DMFT, GA is less computationally intensive and thus allows the description of inhomogeneous systems such as thin films subjected to a perpendicular electric field. 

Although our simplified approach is only qualitative, it gives important information about how one may be able to spatially tune the quasiparticle weight distribution near surfaces and interfaces. This could be relevant for future studies; for example for an inhomogeneous bad metal-superconductor transition by the charge transfer mechanism which may be responsible for the SC-Insulator transition observed at the interface of a band insulator and a strongly correlated material\cite{caviglial}. We will show that by applying a perpendicular electric field, charges will be trapped at the surface of the Mott insulator and shift the MIT for the surface states.

The outline of the paper is as follows. In section II we review the concept of GA and how the inclusion of on site potentials may change the situation. In section III we introduce our model for the slab geometry, present the numerical scheme used and analyze the corresponding results. Finally in section IV we present our conclusions.
\section{Gutzwiller approximation in the presence of an electric field}
In order to address the narrow band effects in transition metals with d or f orbitals for which correlation effects
play a major role in the behavior of the system the simplest model that is able to explain the most important
terms of the Coulomb interaction between electrons is the well known Hubbard model,
\begin{equation}
 \hat{H}_U = -\sum_{<ij>\sigma} t_{ij} (c^\dagger_{i\sigma}c_{j\sigma} + c^\dagger_{j\sigma} c_{i\sigma} ) + \sum_i U \nup i\ndw i .
\end{equation}
We will describe the ground state properties of the Hubbard model by using the Gutzwiller approximation which
 suppresses the contribution of high energy configurations (here configurations with higher number of double occupancies). This is done by introducing a trail wave function which contains variational parameters to be used subsequently to minimize the total energy of the system. Our aim is to investigate the properties of a strongly correlated system in the presence of an external electrical field which will appear in the Hamiltonian as a position dependent potential.
 The induction of such an inhomogeneity is not random and we still have translational invariance in the direction perpendicular to the applied field. To study the ground state properties in the absence of the electric field, the Gutzwiller wave function is defined as:
\begin{equation}
  | \psi_g \rangle = \prod_i g_i^{\D i} | \psi_{0} \rangle = \prod_i \left[1-\left(1-g_i\right)\D i\right] | \psi_{0} \rangle,
\end{equation}
where the double occupancy operator is $\D i = \nup i \ndw i  $, the variational parameters $g_i$ are introduced to reduce the contribution of high energy configuration's 
in the many body wave function $ | \psi_g\rangle$, and $| \psi_{0}\rangle $ is the unprojected non-interacting (Fermi sea) many body wave function. Although it is obvious that by the inclusion of on site potentials
no new variational parameters are needed because they do not induce any new correlations since the term is a 
single body interaction, nevertheless we will prove it rigorously. To obtain the normalization factors in the limit of spatial infinite
dimensions, for which the Gutzwiller approximation is exact \cite{geb,MV2}, we have to remove spatial correlations which 
occur in infinite dimensions together with on-site Hartree contributions which remain in the $d=\infty$ limit. This can be done
by introducing a new expansion parameter following the guidelines of Ref~[\onlinecite{geb}].
If we include on-site potentials for capturing the effects of external fields, the Hamiltonian becomes:
\begin{equation}
 \hat{H} = \hat{H}_U + \sum_{i\sigma}  v_i\hat{n}_{i\sigma}.
\label{eq:H}
\end{equation}
In order to find the ground state of the Hamiltonian in Eq.~(\ref{eq:H}) we introduce new variational parameters, $\zeta_{i\sigma}$ and $\zeta_{i\bar{\sigma}}$, to decrease the weight of the occupancy of the sites with higher on-site energy. The Gutzwiller wave function now becomes:
\begin{eqnarray}
 |\psi_g\rangle &=& \left[1-(1-\zeta_{i\sigma})\nup i\right]\left[1-(1-\zeta_{i\bar{\sigma}})\ndw i\right]\times \nonumber \\ 
  && \left[1-(1-g_i)\D i\right]|\psi_0\rangle
\end{eqnarray}
The standard way of removing on-site Hartree contributions is to introduce the fugacity factors $\muup i$ and $\mudw i$ \cite{geb}, the expansion parameter $x_i$  and the non interacting state $| \varphi_0 \rangle $. Then the Gutzwiller wave-function can be written as: 
\begin{equation}
\begin{split}
&| \psi_g \rangle = \prod_i {\zeta_{i\sigma}}^{\nup i}{\zeta_{i\bar{\sigma}}}^{\ndw i}{g_i}^{2(\gamma_i-\mudw i\ndw i-\muup i \nup i+\D i)} | \varphi_0 \rangle\\ 
&= \prod_i (1+{x_i}(\hat{D}_i -\hat{D_i}^{HF}))| \varphi_0 \rangle.\\
\end{split}
\end{equation}
The Hartree double occupancy operator can be defined as $\hat{D_i}^{HF} = \nup i \ndwzr i +\nupzr i \ndw i - \nupzr i \ndwzr i $ and it is the result of the usual mean field decomposition $\nup i \rightarrow \nup i-\nupzr i$. By defining 
$\zeta_{i\sigma}={g_i}^{\beta_{i\sigma}}$, $\zeta_{i\bar\sigma}={g_i}^{\beta_{i\bar\sigma}}$, 
${\muup i}' =\beta_{i\sigma} + \muup i $ and ${\mudw i}' =\beta_{i\bar\sigma} + \mudw i $ we have:
\begin{equation}
 | \psi_g \rangle = \prod_i {g_i}^{2(\gamma_i-{\mudw i}'\ndw i-{\muup i}' \nup i+\D i)}| \varphi_0 \rangle.
\end{equation}
Therefore by using the above change of variables it is possible to obtain the same renormalization factors for the infinite 
dimensions limit as stated in \cite{geb}. Moreover by using the condition $\langle \nup i \rangle =  \nupzr i $ which holds
for infinite dimensions it can be inferred that the physical counterparts of the new variational parameters, $\zeta_{i\sigma}$, are $\nupzr i$, 
 In minimization procedure we need to minimize the energy with respect to $|\varphi_0\rangle$ together with 
local variational parameters $g_i$ that one needed to describe the correlation effects. In short, the addition of on-site potentials does not
add any new variational parameters and the procedure of finding the ground state is the same as in the conventional
Gutzwiller method. Thus the expectation value of the Hamiltonian: 
\begin{equation}
\begin{split}
\langle \hat{H} \rangle =& -\sum_{ \langle ij \rangle\sigma}\sqrt{\qup i} \sqrt{\qup j} t_{ij}\langle\varphi_0|\cdgiup j\ciup i+ h.c. |\varphi_0\rangle\\
&+\sum_i V_i \langle\varphi_0| \hat{n_i} |\varphi_0\rangle + \sum_i U\bar{d_i}
\end{split}
\end{equation}
has to be minimized only with respect to $g_i$ and $|\varphi_0\rangle$. Here the renormalization factors $\qup i$ depend on the local density of the non-interacting state $|\varphi_0\rangle $ and $g_i$:
\begin{equation}
\begin{split}
 &q_{i\sigma} = \frac{1}{\langle \nup i \rangle_0(1-\langle \nup i \rangle_0)}\times\\
&{\left[{\sqrt{d_i( \langle \nup i \rangle_0- d_i)}+\sqrt{(\langle \ndw i \rangle_0- d_i)(1 - n_{i,0} + d_i)} }\right]}^2,\\
\end{split}
\end{equation}
where $n_{i,0} = \nupzr i + \ndwzr i $, while $g_i$ are described by the following equations which holds in infinite dimensions:
\begin{equation} 
 {g_i}^2 = \frac{d_i(1-n_{i,0}+d_i)}{(\langle \ndw i \rangle_0- d_i)(\langle \nup i \rangle_0- d_i)}
\end{equation}

Although in normal metals any deviation from half filling may lead to the lowering of the electron conductivity, in strongly correlated materials these deviation play different role because of the dependence of renormalization factors of tight binding parameter, $\qup i$, on the local charge density. Thus one may predict that if an applied electrical field would be able to change the charge distribution of the system then it will be able to change the electron conductivity and even shift the metal-insulator transition point. 

In practice minimizing the expectation value of the Hamiltonian is difficult because of the existence of a large number of variational parameters in $|\varphi_0\rangle $ together with the dependence of the renormalization factors on $|\varphi_0\rangle $. This will lead to a highly nonlinear set of equations. In order to alleviate some of the difficulties it is possible to allow local densities and  $|\varphi_0\rangle $ 
to vary independently in the minimization procedure. Then by introducing $\lambda_{i\sigma} $ as Lagrange multipliers it is possible to ensured 
that the local charge densities of the Gutzwiller wave function are equal to the local charge densities of the non interacting state. Other multipliers, $\Lambda$ and $ E $, are introduced in order to ensure total charge conservation and guarantee that $|\varphi_0\rangle $ is normalized.
Therefore the final form of the energy expectation value is:
\begin{eqnarray}\label{htot}
\langle \hat{H} \rangle &&= -\sum_{ \langle ij \rangle\sigma}\overline{t}_{ij}\langle\varphi_0|\cdgiup j\ciup i+ h.c. |\varphi_0\rangle \nonumber\\
&&+\sum_{i,\sigma} v_i \langle\varphi_0| \nup i |\varphi_0\rangle + \sum_i U\bar{d_i}+\sum_{i,\sigma}\lambda_{i\sigma}(\nupzr i - n_{i\sigma}) \nonumber\\
&&+\Lambda(N-\sum_{i\sigma} n_{\sigma i})+E(1-\langle\varphi_0|\varphi_0\rangle),
\end{eqnarray}
where $\overline{t}_{ij}=\sqrt{\qup i} \sqrt{\qup j} t_{ij}$ are the renormalized hopping amplitudes.
To find the optimum energy first of all we vary the  $|\varphi_0\rangle$ for which we have the following Schr\"odinger like equation: 
\begin{equation}\label{mdiag}
\sum_{ \langle ij \rangle\sigma}-\overline{t}_{ij} (\cdgiup j\ciup i+ h.c. )|\varphi_0\rangle 
+\sum_i (V_i+\lambda_{i\sigma})  \nup i |\varphi_0\rangle = E |\varphi_0\rangle,
\end{equation}
which has to be diagonalized for both spins. The sums are up to the filling of the system. This non-interacting
energy is the amount of kinetic energy which is stored in the quasiparticle state $|\varphi_0\rangle$. Then $|\varphi_0\rangle$
is substituted in eq.(\ref{htot}) and the expectation value becomes: 
\begin{equation}
\langle \hat{H} \rangle  =  E_{NI}+ \sum_i U\bar{d_i} +\Lambda(N-\sum_{i,\sigma} n_{i\sigma})+ \sum_{i,\sigma} \lambda_{i\sigma} n_{i\sigma}
\end{equation}
where $E_{NI}$ is the non-interacting energy which depends on the variational parameters $n_{i\sigma}$, $\lambda_{i\sigma}$ and $|\varphi_0\rangle$.
$|\varphi_0\rangle$ is now a function of the variational parameters $\lambda_{i\sigma}$, $n_{i\sigma}$ and $g_i$, and
the above energy functional has to be minimized in accordance to all these parameters. This leads to the following set of 
saddle point conditions:
\begin{equation}\label{nlset}
\begin{split}
 &\frac{\partial \langle \hat{H} \rangle}{\partial \Lambda} = 0, \\
 &\frac{\partial \langle \hat{H} \rangle}{\partial \lambda_{i\sigma}} = 0,\\
 &\frac{\partial \langle \hat{H} \rangle}{\partial n_{i\sigma}} = 0,\\
 &\frac{\partial \langle \hat{H} \rangle}{\partial g_i } = 0.\\
\end{split}
\end{equation}
\section{Model and Numerical scheme}
\subsection{Model}
Our model is a slab geometry in which we have translational invariance in $x$ and $y$ direction and finite size in $z$ direction. In addition we apply a linear potential profile from $-v/2$ to $+v/2$, in the z-direction. With the above assumptions the expectation value of the Hamiltonian can be written as:
\begin{eqnarray}\label{hexpect}
 \langle\hat{H}\rangle &&= \langle\varphi_0|\sum_{i,k_\|,\sigma}(-2tq_{i\sigma}(cosk_x+cosk_y)+v_{i}+\lambda_{i\sigma})\cdgiup{ik_\|}\ciup{ik_\|} \nonumber\\
 &&-\sum_{<ij>k_\|\sigma} \sqrt{\qup i}\sqrt{\qup j}t(\cdgiup{ik_\|}\ciup{jk_\|} + \cdgiup{jk_\|}\ciup{ik_\|})|\varphi_0\rangle\nonumber\\
 &&-N_{k_\|}\sum_{i \sigma}\lambda_{i\sigma}{n}_{i\sigma}+\varLambda(N_{k_\|}\sum_{i\sigma} {n}_{i\sigma} - N ) + E(1-<\varphi_0|\varphi_0>)\nonumber\\
 &&+  \sum_{i} N_{k_\|} U d_i,
\end{eqnarray}
where $i$ and $j$ correspond to atoms in the $z$ direction and $N_{k_\|}=N_{k_x}N_{k_y}$ is the total number of k points. 

First we minimize the energy with respect to $| \varphi_0 \rangle$ which leads to the following eigenvalue problem:
\begin{equation}\label{eigen}
\begin{split}
&\sum_{i,k_\|,\sigma}(-2tq_{i\sigma}(cosk_x+cosk_y)+v_{i}+\lambda_{i\sigma})\cdgiup{ik_\|}\ciup{ik_\|}|\varphi_0> -\\
&\sum_{<ij>k_\| \sigma} \sqrt{\qup i}\sqrt{\qup j}t(\cdgiup{ik_\|}\ciup{jk_\|} + \cdgiup j \ciup i)|\varphi_0>  = E_{k_\|}|\varphi_0>\\
\end{split}
\end{equation}
Eq.~(\ref{eigen}) has to be solved for each $k_\|$ point and in order to find the non-interacting ground state the eigenvalues will be summed up to the desired filling level:
\begin{equation}
 E_{NI}=\sum_{E<E_F} E_{k_\|,n},
\end{equation}
where $E_F$ is the Fermi energy of the quasiparticle states and $n$ is the quantum number for the energy level of each k point.

In the next step the above non-interacting state $|\varphi_0>$, which is now an implicitly function of  all variational parameters $\lambda_{i\sigma}$ , $n_{i\sigma}$, $g_i$ and $\varLambda$, should be inserted into Eq.~(\ref{hexpect}):
\begin{equation}
\begin{split}
& \langle\hat{H}\rangle= <\varphi[\lambda_{i\sigma} , n_{i\sigma} , g_i ,\varLambda]|\hat{H}_0|\varphi[\lambda_{i\sigma} , n_{i\sigma} , g_i ,\varLambda]> \\
&+ N_{k_\|}\sum_{i,\sigma}\lambda_{i\sigma} n_{i\sigma} + \varLambda(N_{k_\|} \sum_{i,\sigma} n_{i\sigma} - N ).
\end{split}
\end{equation}
We therefore minimize the total energy according to the variational parameters by considering:
\begin{equation}
 \frac{\partial}{\partial \lambda}<\psi[\lambda]|H_0[\lambda]|\psi[\lambda]> = <\psi[\lambda]|\frac{\partial}{\partial \lambda}H_0[\lambda]|\psi[\lambda]>,
\end{equation}
which holds when the wave-function is an eigenfunction of the non-interacting Hamiltonian and obtain the following set of saddle point equations for the paramagnetic case ($<\nup i> = <\ndw i>$):
\begin{widetext}
\begin{equation}\label{nlst1}
\begin{split}
&\frac{\partial <\hat{H}>}{\partial g_i} = 2\langle\varphi_0|\sum_{ik_\|} -2t(cosk_x + cosk_y)\frac{\partial \qup i}{\partial g_i} \cdgiup{ik_\|}\ciup{ik_\|}\\
 &-\delta_{i,j\pm1}\sqrt{\frac{\qup j}{\qup i}}\frac{\partial \qup i}{\partial g_i} t(\cdgiup{ik_\|}\ciup{jk_\|} + \cdgiup{jk_\|}\ciup{ik_\|})|\varphi_0> + \sum_{i} N_{k_\|} U\frac{\partial d_i}{\partial g_i}=0,\\
\end{split}
\end{equation}
\begin{equation}\label{nlst2}
\begin{split}
&\frac{\partial <\hat{H}>}{\partial n_{i\sigma}} = 2<\varphi_0|\sum_{ik_\|} -2t(cosk_x + cosk_y)\frac{\partial \qup i}{\partial n_{i\sigma}} \cdgiup{ik_\|}\ciup{ik_\|}\\
&- \delta_{i,j\pm1}\sqrt{\frac{\qup j}{\qup i}}\frac{\partial \qup i}{\partial n_{i\sigma}}t(\cdgiup{ik_\|}\ciup{jk_\|} + \cdgiup{jk_\|}\ciup{ik_\|})\sum_{ik_\|}|\varphi_0>- 2N_{k_\|}\lambda_{i\sigma}- 2N_{k_\|}\varLambda +\sum_{i} N_{k_\|}  U\frac{\partial d_i}{n_{i\sigma}}=0,\\
\end{split}
\end{equation}
\begin{equation}\label{nlst3}
\begin{split} 
&\frac{\partial <\hat{H}>}{\partial \lambda_{i\sigma}} = \langle\varphi_0|\sum_{k_\|} \cdgiup{ik_\|}\ciup{ik_\|} |\varphi_0\rangle - N_{k_\|} n_{i\sigma} = 0, \\
\end{split}
\end{equation}
\begin{equation}\label{nlst4}
\begin{split}
&\frac{\partial <\hat{H}>}{\partial \varLambda} =( N -2N_{k_\|}\sum_{i} n_{\sigma} ) = 0.
\end{split}
\end{equation}
\end{widetext}
In order to numerically solve the above set of non-linear equations, we use MinPack.1\cite{minp1,minp2} which uses a trust-region-dogleg method, while for the k-space summation we choose a $8 \times 8$ Monkhorst-Pack\cite{monkhorst}
k-grid for which the energy is well converged for this kind of grid. From the above equations it is obvious that the Jacobian matrix required by the nonlinear solver has to be calculated by a finite difference method because no analytical evaluation of the Jacobian matrix is possible. Also note that the Jacobian matrix is dense and all of its elements are nonzero.

We also tried to implement another approach by solving Eqs.~(\ref{eigen}) and (\ref{nlst1})-(\ref{nlst4}) iteratively by starting from an estimation of the variational parameters and a calculation of $|\varphi_0>$ which are then supplied to the set of Eqs.~(\ref{nlst1})-(\ref{nlst4}) to find a new set of variational parameters and then repeat the whole procedure. The iterative approach did not converge for values of $U>4t$ which could be because of the high non-linearity of the equations for large $U$. Other authors also reported similar problems with such an iterative scheme \cite{anderas}. 

Although the second approach is less costly, because the Jacobin matrix in the first method is updated at each variation of the parameters it is more likely that the first method converges better particular for large U when we have a large dependence of $|\varphi_0>$ on the variational parameters.
\begin{figure}
\begin{center}
   \includegraphics[width=7cm]{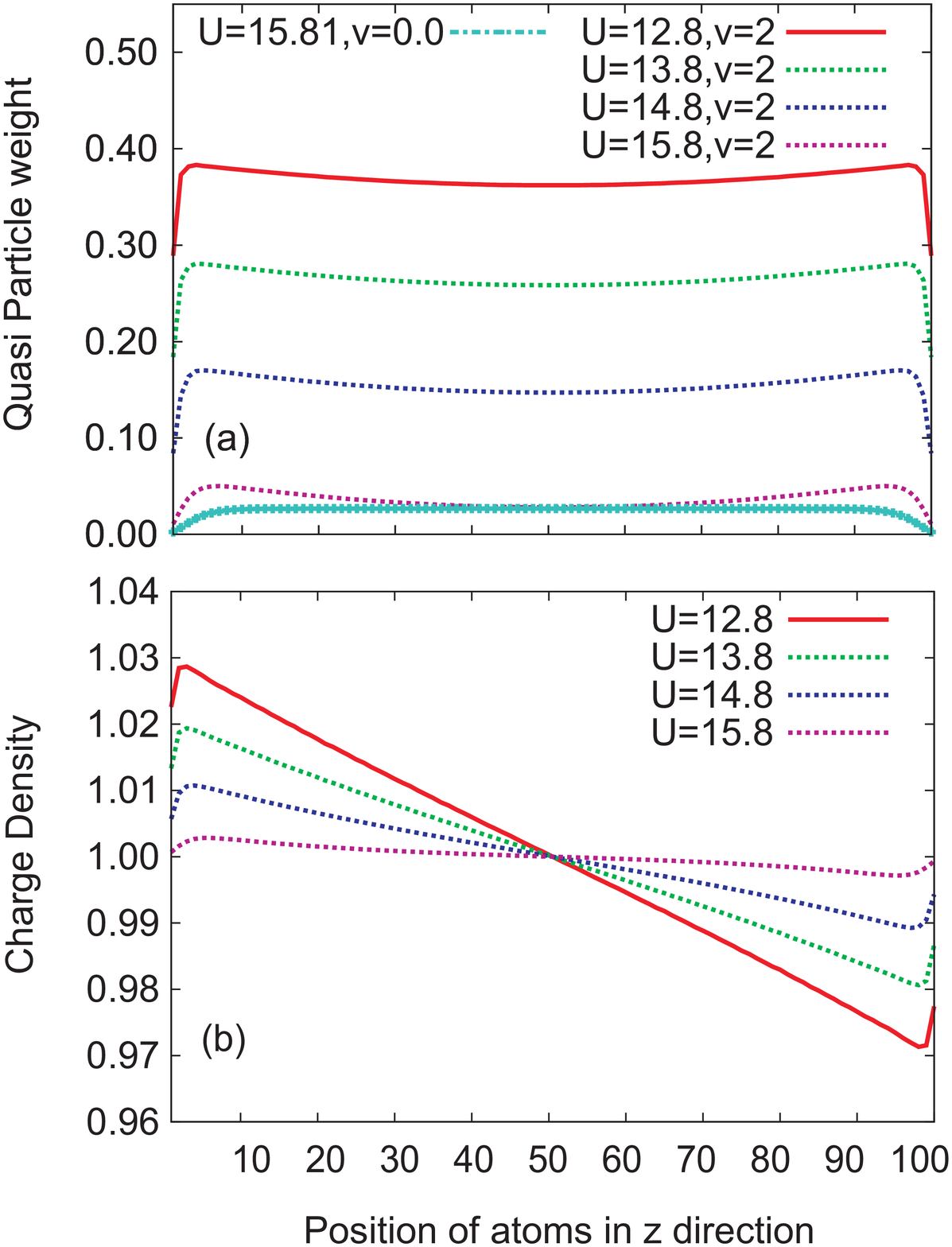}
   \caption {(a) Quasiparticle weight distribution for $U<16t$, $N_z=100$ and $v=2t$ and $v=0$; (b) charge distribution for $U<16t$, $N_z=100$ and $v=2t$. Note that for $v=0$ the system is at half-filling, $n_i=1$.}
   \label{fig1}
\end{center}
\end{figure}
In the next sections we report results for $q_i$ as the position dependent quasiparticle weight which is an indication of the mobility of the electrons in Fermi liquid theory. It is possible to show that the inverse of this factor is proportional to the mass renormalization which is divergent for $q_i=0$ and which corresponds to an insulating phase\cite{fazekas}.
The quantity $\tilde{v_i} = v_i + \lambda_{i\sigma} + \Lambda $ is considered as an effective potential which acts effectively only on $|\varphi_0\rangle$. The parameters $U$ and $v$ are scaled with the tight binding parameter $t$.
\subsection{Numerical results}
We solve the set of Eqs.~(\ref{nlst1})-(\ref{nlst4}) for a slab geometry and a linear distribution of the potential profile in order to investigate its effect on strong correlations. Although we do not consider long range Coulomb interactions or Poisson-Schr\"odinger coupling at this level, it is possible to couple the current solutions to a Poisson solver in order to consider more screening effects.

\begin{figure}
\begin{center}
   \includegraphics[width=7cm]{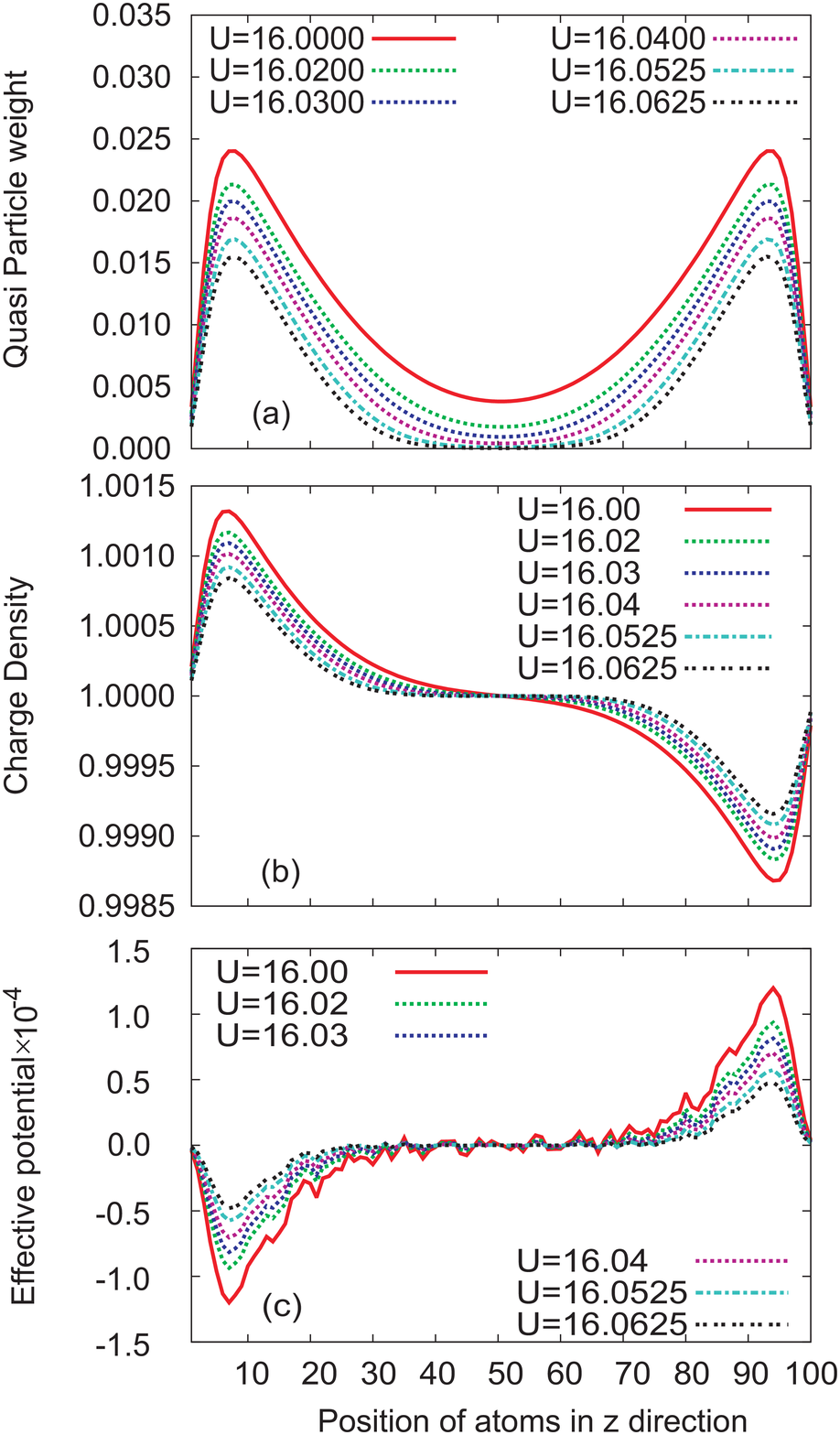}
   \caption {(a) Quasiparticle weight distribution, (b) charge distribution and (c) effective potential for $v=2t$. Notice the $U/2$ contribution is subtracted from the effective potential.}
   \label{fig2}
\end{center}
\end{figure}

The spatial distribution of the quasiparticle weights and the charge densities are shown in Figs.~\ref{fig1} and Fig.~\ref{fig2}(a)-(b) for different values of the Hubbard repulsion $U$ for a slab of width $N_z=100$. Mathematically, the existence of a potential profile causes charge distortion in the system and because of the nature of the Gutzwiller renormalization factors that have a minimum value at half filling($n_i=1.0$) it is predicted that any charge frustration in the system may lead to larger quasiparticle weights when compared to the case without electrical field. 

\begin{figure}
\begin{center}
   \includegraphics[width=7cm]{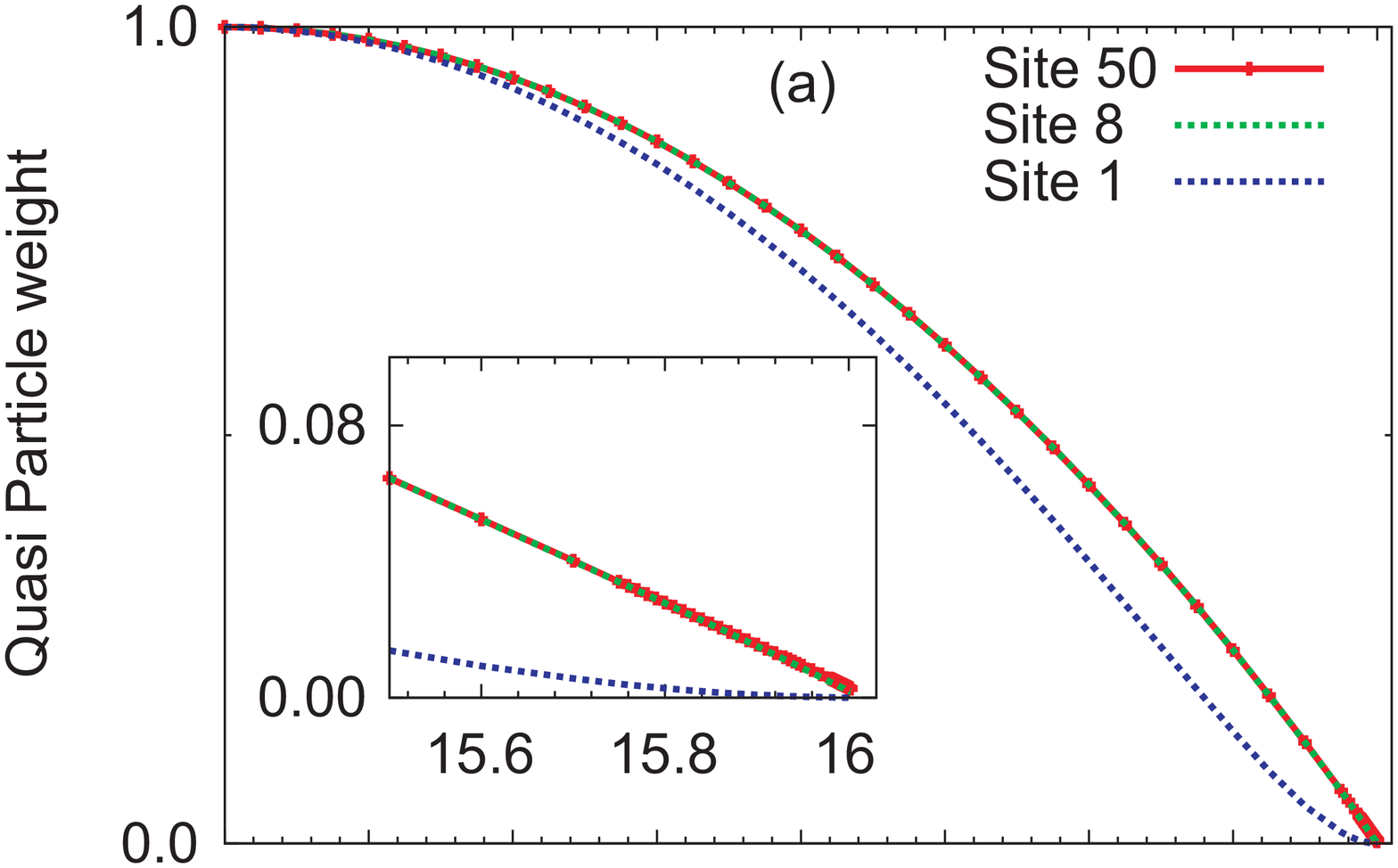}
   \includegraphics[width=7cm]{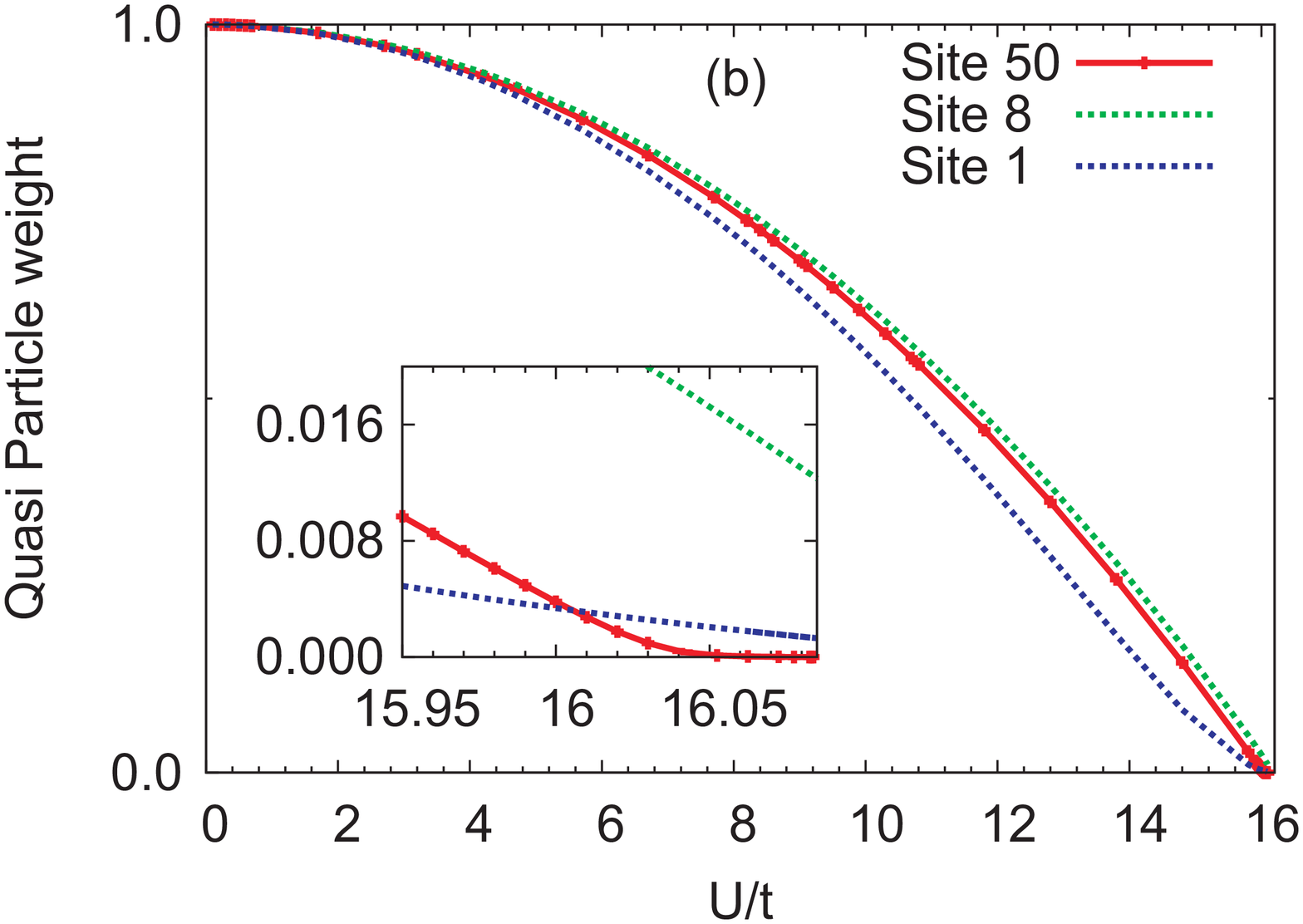}
   \caption {Quasiparticle weight of various sites versus Hubbard repulsion for $N_z=100$, (a) $v=0$ and (b) $v=2t$. }
   \label{fig3}
\end{center}
\end{figure}

For both $U<U_c$ and $U>U_c$ (where $U_c=16t$ for bulk) the maximum quasiparticle weight is achieved in few layers beneath the surface as is obvious from Figs.~\ref{fig1}(a) and ~\ref{fig2}(a). For $U<U_c$ akin to the zero electric field case\cite{nourafkan} the minimum quasiparticle weight is achieved for the surface sites. In contrast, for $U>U_c$ the quasiparticle weight of the central atoms dramatically starts to drop to extremely low values and creates a dead insulating region as is indicated in Fig.~\ref{fig2}(a). This is presented more clearly in Figs.~\ref{fig3}(a)-(b) where we show the quasiparticle weight versus the Hubbard repulsion for three significant locations (surface, near surface and bulk) for both $v=0$ and $v=2t$.

%
\begin{figure}
\begin{center}
   \includegraphics[width=7cm]{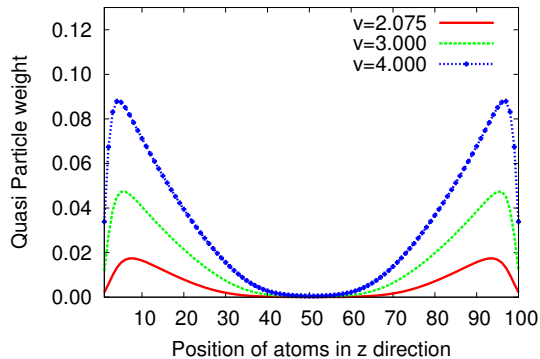}
   \caption {Quasiparticle weight distribution versus $v$ for $U=16.0625t$ and $N_z=100$.}
   \label{fig4}
\end{center}
\end{figure}
The formation of this dead zone together with the fact that we increased the value of the Hubbard repulsion from lower values may lead to charge being trapped near the surfaces of the slab because electrons are not able to tunnel through the bulk. 
This charge trapping prevents the system to exhibit a complete metal-insulator transition even for values of the Hubbard repulsion larger than the bulk $U_c$. 
This result is contrary to the case without electrical field where there is a single $U=U_c$ at which the quasiparticle weight is suppressed for the hole system. 
When there is no electrical field the quasiparticle weight is maximal in central parts as shown in Fig.~\ref{fig1}(a). The distance over which the quasiparticle weight 
recovers its bulk value is on the order of 10 atoms which shows that the electrons which are located on the surface atoms suffering from the lack of kinetic energy 
(due to lower coordination number at the surface) are always able to gain kinetic energy from the central sites with highest quasiparticle weight. Thus the surface 
quasiparticle weights will not vanish completely as long as the bulk quasiparticle weight is finite, although may have
very low values.\cite{potthof,nourafkan,michele} 

To see the difference between cases with $v=0$ and $v\neq0$ it should be noticed that in the case in which electrical field is present the quasiparticle weight is maximal in few layers beneath the surface. 
This is because of the fact that electrons at these locations are more intensely affected by the electrical field while they do not suffer from the lack of kinetic energy as do the electrons corresponding 
to cites which are exactly at the surface. Therefore the distance between the maximum quasiparticle weight (as source of kinetic energy) and central sites is higher specially for higher sizes and as a result 
the central sites are not able to gain kinetic energy, moreover this sites are less affected by electrical field when one increases the width of the slab together with fixing potential difference between edges
to a constant value as we have consider in our model. Thus the metal-insulator transition occur for sufficiently large width and some $U>U_c$ for central sites before complete screening of electrical field
as indicated in fig.~\ref{fig3}(b) more precisely. This is very similar to the case of the interface of a bad metal with a strongly correlated material with $U>U_c$, 
where there is an insulating phase sufficiently far from the source of kinetic energy which is located near the surface.\cite{costi,michele}

In Figs.~\ref{fig2}(b) and ~\ref{fig2}(c) the spatial distribution of the charge densities and the effective potentials are shown for different values of $U>U_c$. 
Both of these two quantities behave similarly to the quasiparticle weight. The charge density is maximum in the same location in which we have the maximum quasiparticle weight while for the sites with charge density near local half filling ($n_i=1.0$) we have the lowest quasiparticle weight and this is where the electrical field has the weakest effect. 
This confirms that the higher quasiparticle weight is due to a larger carrier density near the surface of the slab. The deviations of the carrier densities from half-filling correspond to larger electron density for sites with lower effective potential and hole density for sites for higher effective potential as shown in Fig.~\ref{fig2}(c). 
The charge frustration is responsible for nonzero quasiparticle weight for these sites near the surfaces of the system even for $U>U_c$. 
The regime of nonzero conductivity of the edge regions for $U>U_c$ is similar to underdoped and overdoped Mott insulators but in this case we have an inhomogeneous charge distribution due to the presence of the electrical field.

\begin{figure}
\begin{center}
   \includegraphics[width=7cm]{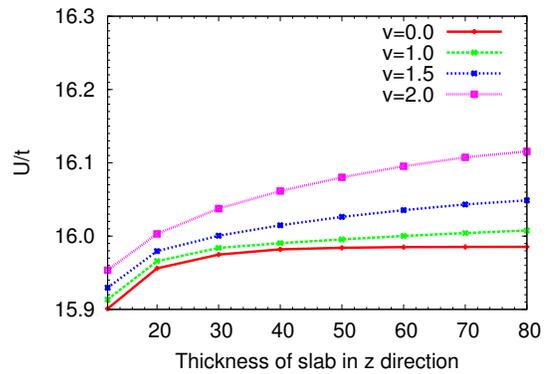}
   \caption {Critical Hubbard repulsion for which the maximal quasiparticle weight is $Z=5.0\times10^{-3}$ versus slab thickness for different electric fields.}
   \label{fig5}
\end{center}
\end{figure}

Fig.~\ref{fig4} shows the change of quasiparticle weight throughout the system when the voltage difference is increased. While the location of the maximal quasiparticle weight slowly shifted towards the surface, its value increases with electric field. This in turn assures that the size of the central dead zone reduces when the voltage difference is increased. One should note that when measuring an I-V curve only in-plane the conductivity will show metallic behavior because the z-axis conductivity will be dominated by the bulk insulating layer.

\begin{figure}
\begin{center}
   \includegraphics[width=7cm]{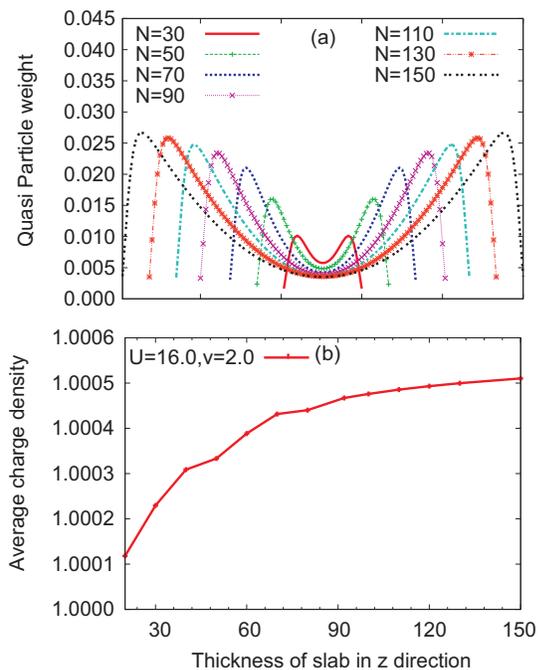}
   \caption {(a) Quasiparticle weight distribution of different sites for various  slab thicknesses; (b) The charge density averaged over half of the slab for different thicknesses. Here $U=16t$ and $v=2t$.}
   \label{fig6}
\end{center}
\end{figure}
Fig.~\ref{fig5} shows the value of the Hubbard repulsion for which the maximum quasiparticle weight is $Z=5.0\times10^{-3}$ as a function of slab thickness. This will give a lower bound for the critical $U_c^{slab}$ in the presence of a perpendicular electric field. $U_c^{slab}$ is higher for larger thicknesses and stronger fields $v$. Again this is related to the amount of charges localized near the surfaces. When $U$ increases the quasiparticles corresponding to central parts drop faster in the thicker slabs as is indicated in Fig.~\ref{fig6}(a) and this causes more charge accumulation at the surfaces. This is because of the fact that the probability of the electrons to tunnel through the central parts is being reduced which makes charge relaxation more difficult. In other words by increasing the Hubbard repulsion the system tries to screen the charges due to energy restrictions while on the other hand this increasing of U suppress metalic behavior of central part and thus hinders the charge relaxation. This scenario is expected to be even more relevant for thicker slabs. This can be better understood  by considering the average charge accumulation in half of the slab which increases for thicker slabs as indicated in Fig.~\ref{fig6}(b). 
\begin{figure}
\begin{center}
   \includegraphics[width=7cm]{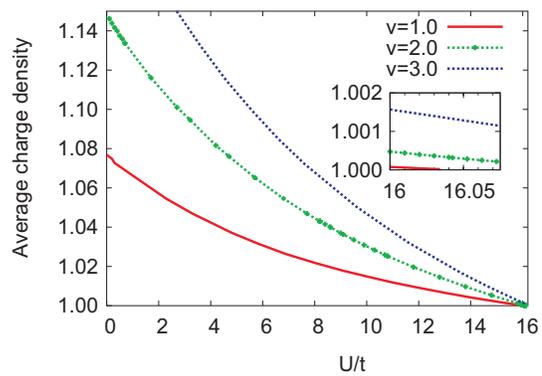}
   \caption {The average charge density for one side of slab for $N_z=100$.}
   \label{fig7}
\end{center}
\end{figure}

Fig.~(\ref{fig7}) shows the charge accumulation in one side of the slab as a function of the Hubbard repulsion for different strengths of the electric field. While the charge accumulation decreases for higher Hubbard repulsion due to screening, it instead increases for higher voltage values.
 Although we did not obtain a clear asymptotic behavior for maximum quasiparticle weight by increasing the slab width, it seems that one may gets an asymptotic solution for very thick slabs, in which the central sites may have a MIT at $U$ very close to $16t$. This may occur due to the very 
large distance of the central sites from the edges. There are two reasons that the central sites yield MIT very near $16t$ for very thick slabs: first of all the central parts have the lowest charge density deviation from $n=1.0$ because of the shape of the potential profile, second because
 it is difficult for these sites to gain kinetic energy from the source of kinetic energy which is located in only few layers beneath the surfaces. By considering the latter facts an asymptotic solution may be achieved for extremely thick slabs.

\section{Conclusions}
In conclusion, we described the Mott metal-insulator in a slab geometry in the presence of an external electrical field by calculating the site dependent quasiparticle weight. This is done by using an inhomogeneous Gutzwiller approximation which is exact in the limit of infinite dimensions. 
Increasing the Hubbard repulsion from lower values in the presence of an external electrical field leads to the formation of a dead insulating zone
at the center of the thin film. The formation of the dead zone for $U>16t$ happens before the complete screening of the electrical field and
therefore charge trapping occurs at the edge sites. This charge trapping causes the MIT to be shifted for edge sites in the presence of the external field. We therefore show that even though the central region becomes insulating at $U_c$, the surface layers remain metallic but with a suppressed
 quasiparticle weight. From an experimental point of view our results are relevant for transport measurements in thin films. In the presence of an external electric field perpendicular to an insulating film, one could use the surface states for transport since the charge transfer at the surface creates two dimensional underdoped/overdoped regions. In the same time, transport perpendicular to the thin film is suppressed due to the dead insulating zone, thus protecting the surface states from leakages. The electric field needed to create the surface states is also much lower than the breakdown field needed to pass current across the insulating zone.

\begin{acknowledgements}
 This work was supported by the Flemish Science Foundation (FWO-Vlaanderen) and the Belgian Science Policy (IAP). L.C. acknowledges individual support from FWO-Vlaanderen.
\end{acknowledgements}

\end{document}